**Emergence of physical properties mapped in a two-component system**

Jason N. Armstrong¶, Eric M. Gande¶, John W. Vinti, Susan Z. Hua and Harsh Deep Chopra*

*Laboratory for Quantum Devices, Materials Program, Mechanical and Aerospace Engineering Department, The State University of New York at Buffalo, Buffalo, NY 14260, USA*

Emergence of various properties is mapped in a two-component system from quantum to mesoscale, using Au-Ag alloys. Experiments are designed so that composition is the primary 'interaction' guiding the evolution of properties across 'emergent length scales'. In bulk, Au-Ag form near-ideal solid solutions at all compositions. In contrast, dramatic composition-dependent fluctuations are observed during emergence of atomic cohesion, force required to rupture the bonds, and deformation modes. These fluctuations rapidly smoothen out as volume effects drown out the boundary effects.

Since an atom is the smallest unit that preserves the chemical identity of an element, divergence in properties of different elements originate at the atomic level. Therefore, study of emergent phenomena necessarily entail understanding the evolution of physical properties as atoms assemble together to form larger aggregates.[1-3] For example, size-invariant properties of the bulk become size-dependent at a length scale $L_S$ where surface effects dominate. A simple plot of surface-to-volume ratio versus sample size illustrates the various length scales of interest in the present study. Figure 1(a) shows that regardless of the sample morphology (sphere, cylinder, etc.), boundary effects become noticeable below ~25-35 nm, although they become prominent only below ~3-4 nm. This is validated by recent results on pure Au where crossover from surface to volume dominated deformation was shown to occur at sample diameters between ~2 and 3.2 nm.[4] Below ~1 nm another length scale $L_C$ appears when the number of atoms in the sample becomes comparable to or less than the bulk atomic coordination number (12 for FCC metals like Au and Ag). Bond stiffening has been predicted to occur at this length scale,[5-8] which causes a rapid increase in the forces required to rupture the bonds (strength).[9] Finally, when the sample size approaches the Fermi wavelength of electrons $L_F$ (~0.5 nm in Au and Ag), boundary and atomic coordination effects become even more acute, as illustrated by the size-dependence of strength in the inset of Fig. 1(a).[10] In the following, $L_S$, $L_C$, and $L_F$ are collectively referred to as the 'emergent length scales', i.e., from Fermi wavelength to mesoscale (~3 nm).

In addition to size, composition is another degree of freedom that can be used to alter the bond length, coordination chemistry, electronic structure, etc., and thus alter the pathway of an emergent property at a given temperature. However to-date, its role at the emergent length scales has not been systematically investigated given the considerable experimental challenges to probe the properties of atomic sized samples. The present study investigates the composition dependence of various physical properties across these length scales, using Au-Ag alloys. The choice of Au-Ag alloys allows experiments to be designed so that the role of composition in boundary phenomena is the primary 'interaction' governing the eventual emergence of bulk properties; the choice of Au-Ag is not meant to be a straightforward extension from pure metals to alloys. This is because these two noble metals have same crystal structure (FCC) and valence electrons, with near identical inter-atomic distance (Au: 0.2884 nm; Ag: 0.2884).[11] These similarities make Au-Ag a model system for Hume-Rothery rules governing the extent of solubility in each other;[12, 13] in bulk, Au and Ag are completely soluble in each other at all compositions, forming near ideal solid solutions. However, Ag differs from Au in an important respect. The surface energy of Ag (1200 mJ/m$^2$) is much lower than that of Au (1500 mJ/m$^2$).[14,15]

This difference (along with aforesaid similarities) makes the Au-Ag alloys a model system to study the effect of composition at length scales where surface effects dominate.

In the present study, emergent maps (eMaps) were built as a function of composition and sample size. The approach to build them is reminiscent of how phase diagrams were originally constructed, which required extensive experiments to determine various phases present at a given temperature and composition. Just as the theoretical advances have greatly minimized the painstaking number of experiments involved in building the phase diagrams, one anticipates a similar trend for eMaps given that simulations of atomistic processes in recent years have become highly realistic.[16-19] In the interim, eMaps should help serve as benchmarks for validating theoretical models. The experimental details are described in detail in the supplementary documents.

High cohesive forces between atoms equate with large forces required to rupture the bonds (strength). With pico-level resolution in applied forces and deformation we are able to reduce the samples size atom-by-atom and gain quantitative information on the cohesive forces between the atoms. A typical trace showing such an atom-by-atom deformation process is shown in Fig. 1(b); it also plots the simultaneously measured conductance across the sample that is used to calculate the cross-section area of samples, as described in the section on experimental details. From thousands of such experiments from alloys of different compositions, the eMap for strength was constructed, as shown in Fig. 2; the contour plot of the eMaps in Fig. 2 marks the position of various length scales.[20] A salient feature of Fig. 2 is the steep increase in strength on approaching the Fermi length scale at any given composition. These strength levels are over four orders of magnitude higher than in bulk, and consistent with our recent results on pure Ag and Au, where it was shown that the strength of (defect-free) metals approaches ideal (theoretical) values. Also note that the Au-rich compositions show much higher strength levels at the Fermi length scale than their Ag-rich counterparts. This is consistent with the fact that the observed deformation corresponds to surface modes and Au has higher surface energy than Ag. Practically it means that the strength levels observed in pure silver at any length scale can be achieved by using a much larger diameter sample made of an alternate composition. Another interesting feature of the eMap is the non-monotonic change in strength on crisscrossing from pure Au to pure Ag for a given sample size. It shows that although these alloys form near ideal solutions in the bulk, they are far from following a simple rule of mixtures at the emergent length scales. This behavior is better explained with the help of the elastic constant eMaps in Fig. 3 below.

Zener first pointed out the connection between elastic constants and solubility in bulk alloys.[21] Zener made this connection by noting that misfit strain of solute atoms forces the adjacent atoms to shift to new positions where the radius of curvature of the energy versus position is less sharp. The elastic constants, being proportional to this curvature, are thus reduced. Therefore as a rule, solutes that cause the greatest reduction in elastic constants have the least solubility (discussed later within the context of Hume-Rothery rules). In the present study, misfit between Au and Ag atoms is non-existent due to their virtually identical atomic diameters. Instead, the 'misfit' factor or the dominant interaction that affect the position of atoms at these length scales is the difference in surface energy between Au and Ag, and its effect can be seen from the elastic constant eMap in Fig. 3. It shows that unlike their bulk counterparts, the alloys at these length scales are far from obeying a simple rule of mixtures. Figure 3 shows that in general, the elastic constant for a given composition increases sharply on approaching the Fermi length scale, except for the case of Ag-rich alloys. Also notice that in contrast to Fig. 2, the global maximum for modulus (red) is observed for pure Ag instead of Au. Moreover, this maximum coincides with sharp transition from surface to volume dominated behavior at ~2 nm, instead of the anticipated maximum at the Fermi length scale. We have recently shown that pure silver exhibits four distinct elastic modes at these length scales.[22] Notably, one of the modes stiffens significantly at this size, being 7-8 times higher than that for bulk silver. The observed maximum in modulus for pure silver in Fig. 3 coincides with the stiffening of this mode. In general, the strength is proportional to elastic constant, as predicted by various theories of strength and fracture. However, the difference in position of global maximum in strength and elastic constant eMaps in Fig. 2 and 3, respectively, shows that surface modes behave differently and deviate markedly from classical behavior. Figure 3 also shows a broad minimum in modulus centered around the equiatomic ($Ag_{50}Ag_{50}$) composition that spans all length scales; at the Fermi length scale, the modulus values at these compositions are ½ to ⅓$^{rd}$ of pure gold. It coincides with the broad minimum in strength between 10-75at% Ag.

Finally, Fig. 4 shows the eMap for deformation modes, plotted in terms of the magnitude of discrete atomic displacements δL as sample of a given size is ruptured to another size. It shows that in general, the near-equiatomic Au-Ag alloys undergo larger atomic displacements (red/orange region in the eMap). Previously we have shown that for pure gold, δL transitions from homogeneous shear to defect mediated deformation at ~1.5 nm,[4] as marked in the contour plot in Fig. 2. As a function of alloy composition, this transition is marked by the dark yellow band in Fig. 4, which shows that Ag-rich alloys make this transition at much smaller sizes than gold. We also previously showed that deformation mode in pure Au transitions from being surface dominated to volume dominated at diameters between 2.25 to 3.2 nm;[4] this is also marked in the contour plot in Fig. 2. The composition dependence of this transition is shown by the light yellow band in Fig. 4 and again shows that the Ag-rich alloys undergo this transition at smaller sample diameters. The explanation for this behavior is rooted in a higher surface energy for Au compared to Ag. Imagine a gold sample of a given morphology (say a cylinder) that is being reduced in size in a self-similar manner. Its overall energy would begin to have progressively higher contributions from its surfaces at larger sizes. In contrast, Ag and Ag-rich samples would have lower energy for the same sized sample, and would begin to show pronounced contributions from surface energy at smaller sizes. This is reflected in the data shown in Fig. 4.

Recall that Hume-Rothery rules are used to assess the solubility of elements in each other to form substitutional solid solutions;[12] Hägg described the corresponding rules for solubility in interstitial alloys.[23] Briefly, the four Hume-Rothery rules (atomic size factor, isomorphous crystal structure, similar electromotive potential, and isovalence) dictate the extent of solubility between the alloying elements. Although based on empirical observations, these rules have their origins in the fact that the atomic volumes of metals in substitutional alloys generally tend to be of the same order of magnitude.[13] Results of the present study shows that at the emergent length scales (where surface and atomic coordination effects dominate), the rule guiding the interaction between alloying elements is the difference in surface energy. In hindsight, this is not surprising given the fact that surfaces, and thus surface energy, dominates at these length scales.

As a tool for alloy design, the emergent maps reveal regions of enhanced properties at the mesoscale, obviating the need for extremely small (atomic-sized) architectures that lack long-term stability and are often difficult to implement in practice.

**Acknowledgments:** This work was supported by the National Science Foundation, Grant Nos. DMR-0706074, and DMR-0964830, and this support is gratefully acknowledged.

**FIGURE CAPTIONS**

**Figure 1. (a)** Various length scales that become relevant in study of emergent properties. The plot of surface-to-volume ratio for cylindrical geometry assumes L>>D, or in the present case, assumes that the end caps of the cylinder are excluded since the sample is confined between the AFM tip and the substrate. Upper inset shows size dependence of strength in $Au_{75}Ag_{25}$ alloy. (b) Simultaneously measured force and conductance as an initially large sample is being controllably ruptured down from bulk down to a single-atom. Inset shows a zoom-in view of a section of the force and conductance traces.

**Figure 2.** Emergent map of strength in Au-Ag alloys at room temperature (upper plot) along with its contour representation (below). Various length scales and transition in deformation modes correspond to the case of pure Au.

**Figure 3.** Emergent map of elastic constants in Au-Ag alloys at room temperature.

**Figure 4.** Emergent map of deformation modes in Au-Ag alloys at room temperature plotted in terms of discrete atomic displacements as a sample is ruptured from one size to another.

**References:**

*Corresponding author. Electronic mail: hchopra@buffalo.edu

¶These authors contributed equally to this work.

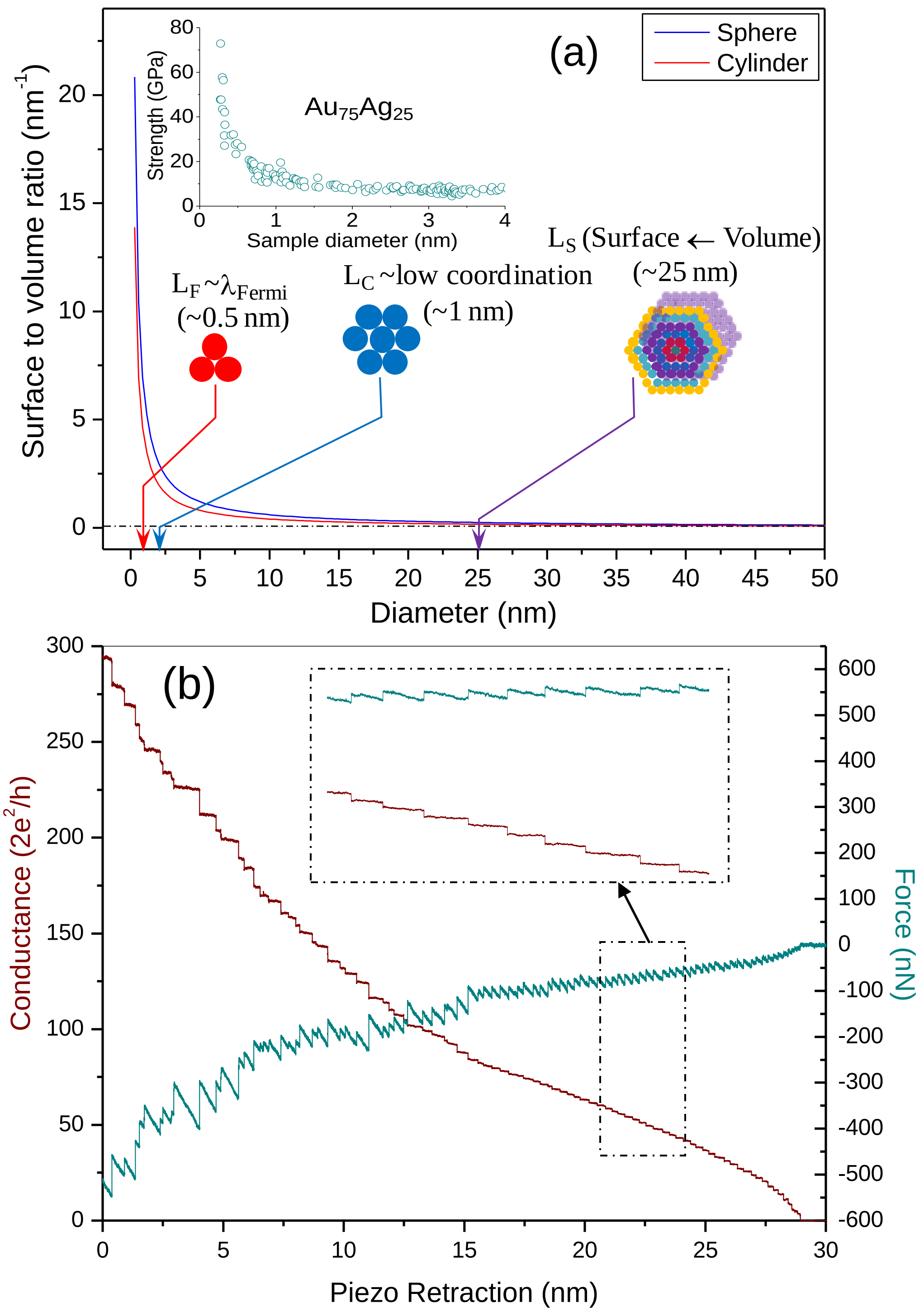

**Figure 1**

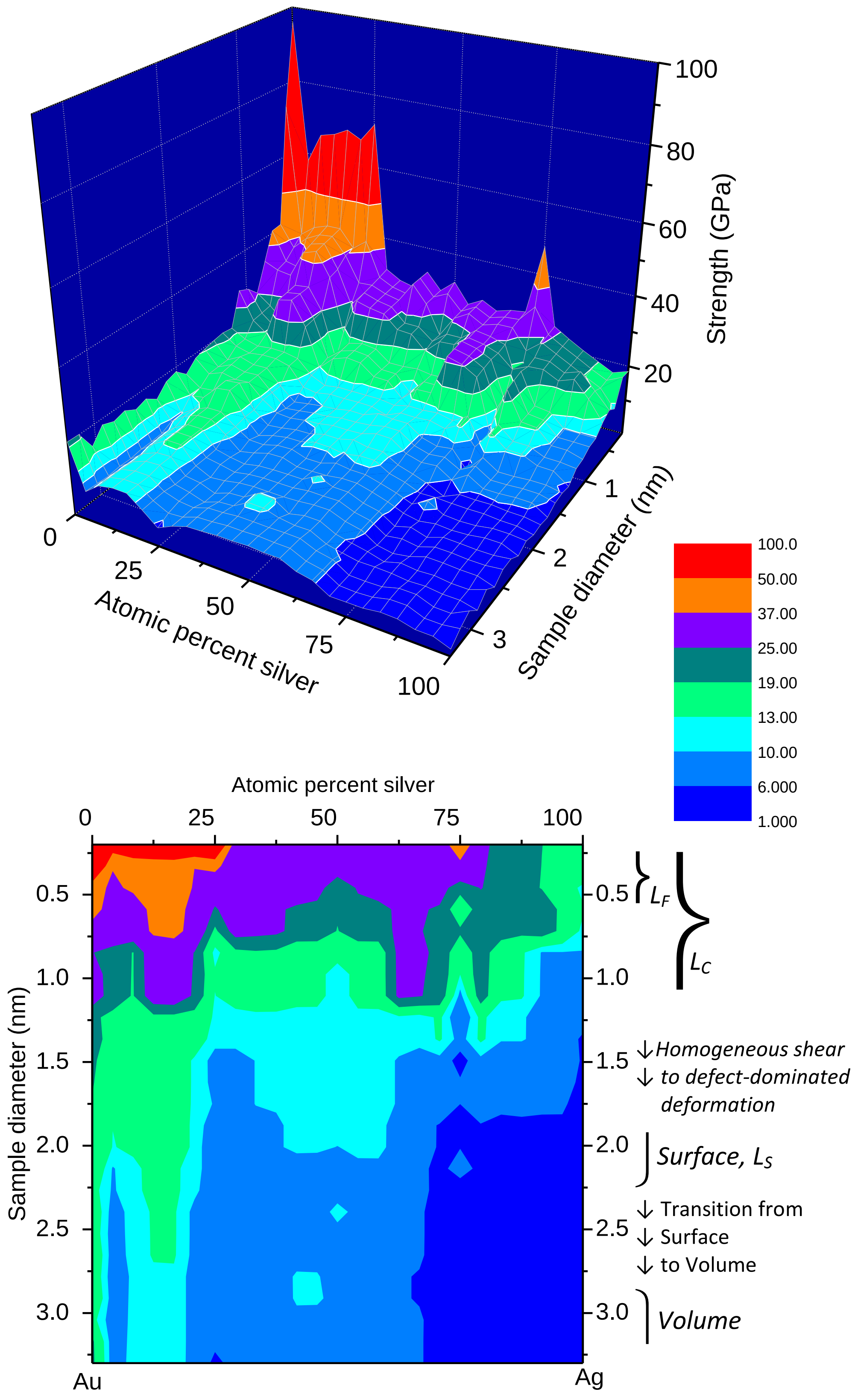

**Figure 2**

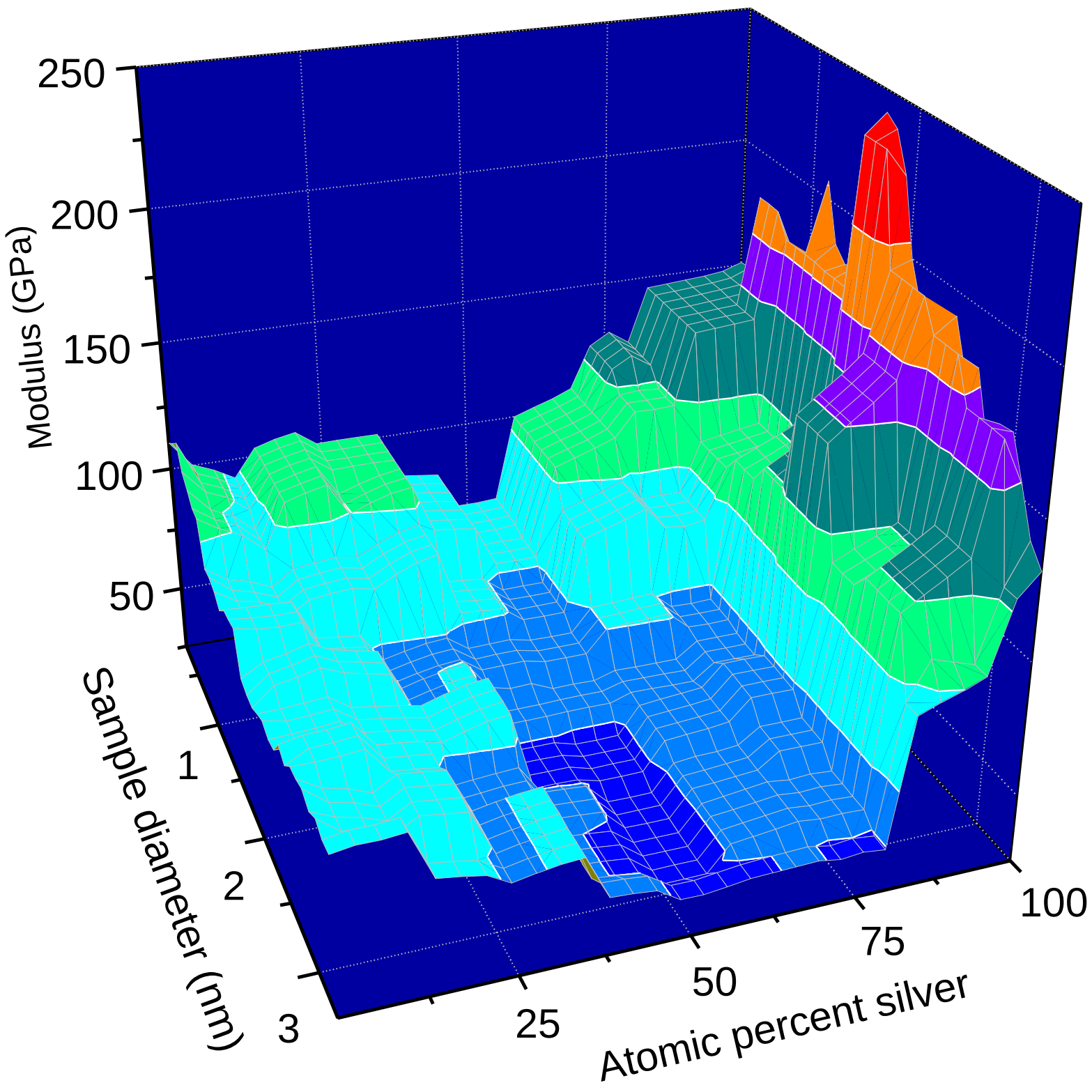

**Figure 3**

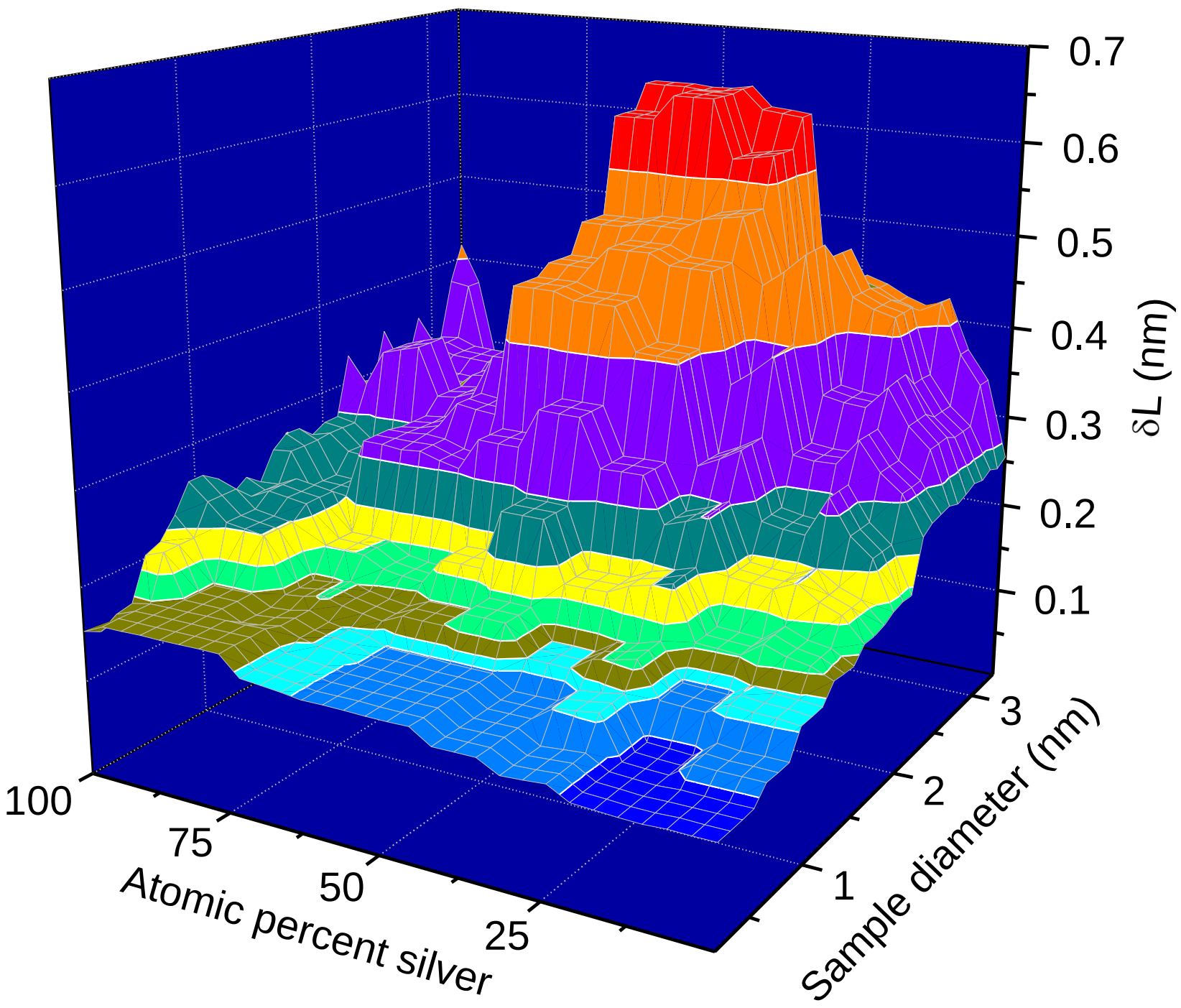

**Figure 4**

**Supplementary Document: Emergence of physical properties mapped in a two-component system**

Jason N. Armstrong, Eric M. Gande, John W. Vinti, Susan Z. Hua and Harsh Deep Chopra

**Experimental details:** Thin films of Au-Ag alloys were made by co-sputtering Au and Ag targets of 99.999% purity. Prior to the deposition, the deposition rate versus sputtering power for each target was calibrated. Post deposition, the composition of each alloy was determined using EDAX. Films (200 nm thick) were deposited on silicon substrates in an argon partial pressure of 3 mtorr in an ultra high vacuum chamber whose base pressure is ~$10^{-8}$-$10^{-9}$ torr. Simultaneously, atomic force microscope (AFM) silicon cantilever tips were also sputter coated with the Au-Ag alloy for force-deformation measurements. During deposition the cantilevers were periodically rotated relative to the sputtering gun to enhance the uniformity of the films. The experimental approach to form and study the mechanics of such samples is described in detail elsewhere.[1, 2] Briefly, a modified AFM (Ambios Q-Scope Nomad) was used for simultaneous measurements of force-deformation and conductance at room temperature in inert atmosphere. The AFM assembly consists of a dual piezo. With this configuration, the noise band is 5 pm (peak-to-peak), and its center line can be shifted by a minimum step of 4 pm. A range of cantilever spring constants was used (16-80 N/m) to study samples as small as a single-atom bridge. The cantilevers were precisely calibrated using reference cantilevers available from Veeco Probes (Force Calibration Cantilevers CLFC-NOBO). The photo-detector was calibrated using the well established optical deflection technique. In all experiments, the piezo was extended or retracted at a rate of 5 nm/s. Simultaneously measured force and conductance across atomic sized silver samples were obtained by breaking an initially large diameter sample down to atomic sized bridges through retraction of the substrate relative to the cantilever tip. Since the spring constant of the cantilever is precisely known, the spring constant of different sized atomic configurations can be calculated by measuring the respective slopes, using the procedure described previously.[1] Our analysis used the simultaneously measured conductance across the samples to calculate their cross-section area using Sharvin formula,[3] with conductance traces recorded at a bias voltage of 50 mV. Briefly, Sharvin formula relates the conductance $G_s$ to the area $A$ of the sample by the relationship $G_s = (2e^2/h)(\pi A/\lambda_F^2) = G_o(\pi A/\lambda_F^2)$; here $(2e^2/h) = G_o$ is the quantum of conductance; $e$ is the quantum of charge; $h$ is Planck constant, and $\lambda_F$ is the Fermi wavelength. At the Fermi length scale, sample diameters can be directly estimated from known values of atomic diameters. However, they differ by less than 5% from the Sharvin analysis and hence all sample sizes were estimated using the Sharvin analysis.